# Plant-Protein-Enabled Biodegradable Triboelectric Nanogenerator for Sustainable Agriculture

Chengmei Jiang[a], Qi Zhang[a], Chengxin He[b], Chi Zhang[a], Xiaohui Feng[c], Xunjia Li[a], Qiang Zhao[b], Yibin Ying[a], and Jianfeng Ping[a,*]

[a] *School of Biosystems Engineering and Food Science, Zhejiang University, 866 Yuhangtang Road, Hangzhou 310058, China*

[b] *State Key Laboratory of Food Science and Technology, Nanchang University, Jiangxi 330047, China*

[c] *Hangzhou Thunder Agricultural Technology Co., Ltd, Hangzhou 311100, China*

*\* Corresponding author: Prof. Jianfeng Ping  jfping@zju.edu.cn*

**Abstract**

As the use of triboelectric nanogenerators (TENGs) increases, the generation of related electronic waste has been a major challenge. Therefore, the development of environmentally friendly, biodegradable, and low-cost TENGs must be prioritized. Having discovered that plant proteins, by-products of grain processing, possess excellent triboelectric properties, we explore these properties by evaluating the protein structure. The proteins are recycled to fabricate triboelectric layers, and the triboelectric series according to electrical properties is determined for the first time. Using a special structure design, we construct a plant-protein-enabled biodegradable TENG by integrating a polylactic acid film, which is used as a new type of mulch film to construct a growth-promoting system that generates space electric fields for agriculture. Thus, from the plant protein to the crop, a sustainable recycling loop is implemented. Using bean seedlings as a model to confirm the feasibility of the mulch film, we further use it in the cultivation of greenhouse vegetables. Experimental results demonstrate the applicability of the proposed plant-protein-enabled biodegradable TENG in sustainable agriculture.

***Keywords***: energy harvesting; triboelectric nanogenerator; plant protein; sustainability, biodegradable; nanodevice

# 1. Introduction

With the widespread development of portable and flexible electronics, triboelectric nanogenerators (TENGs), which can convert mechanical energy from diverse sources into electrical energy, are expected to supply substantial energy in the future [1-3]. To date, most studies on TENGs have emphasized their designs or materials. However, sustainable materials or devices are rarely used [4], and widely used TENG materials, including polytetrafluoroethylene (PTFE) [5, 6], fluorinated ethylene propylene [7], polydimethylsiloxane (PDMS) [2, 8, 9], and polyethylene terephthalate (PET) [10-12], do not degrade completely and eventually release hazardous substances. Consequently, such TENGs will likely become electronic waste with the rapid and continuous upgrading of electronics, imposing a substantial burden on the environment [4, 13]. Therefore, biodegradable and environmentally friendly materials must be adopted in TENGs [14]. Biomaterials such as starch [4], chitosan [15], silk nanofiber [16, 17], gelatin [18], and bacterial nanocellulose [19] have been employed as dielectrics for biomaterial-based biodegradable TENGs (bio-TENGs); however, the current development of bio-TENGs is at an early stage, and the inherent deficiencies of low output power density and monotonous triboelectric behavior should be addressed. Moreover, there is a paucity of research on the operation of biomaterial-based bio-TENGs, greatly restricting their large-scale application. Therefore, exploring more biomaterials and their operation mechanisms has become increasingly important to overcome current TENG limitations.

Plant proteins are a particularly interesting alternative to synthetic polymers given their variety and relatively low price due to their availability as agricultural by-products [20, 21]. Owing to the harsh extraction conditions, these proteins do not compete in the food industry and can be used as biomaterials. Additionally, the excellent film-forming capabilities, abundant nitrogen-containing groups, and diversity of polymerization reactions provide unique opportunities for these proteins [21, 22]. Thus, plant proteins may be promising materials for bio-TENGs. Unlike animal proteins, plant proteins are more affordable and can be obtained from more sources. Nevertheless, they remain underused and have limited application, despite possessing great potential for application in bio-TENG dielectrics.

In recent years, many studies have shown that high voltage can promote plant photosynthesis and respiration and increase metabolism, thus facilitating plant growth and development [23-25]. Additionally, high voltages can ionize negative oxygen ions, playing an important role in air disinfection and purification. Compared with traditional agriculture, the space electric field produces no pollution or residue and has a low energy consumption while promoting crop growth. Accordingly, research on high-voltage electrostatic equipment for agricultural production is being actively conducted [25, 26]. Being a high-voltage and low-current output device, a TENG can produce space electric fields for agriculture, which may provide an alternative for high-voltage electrostatic equipment. Film mulching, another conventional cultivation technique, can greatly increase soil moisture and improve temperature conditions, prevent weed growth, and create a favorable ecological environment for crop growth to enhance crop production [27]. Thus, it is important in improving grain and economic crop yields worldwide.

In this study, we recycled plant proteins and used them as triboelectric materials for the first time. In addition, we analyzed the triboelectric charging behaviors of diverse representative plant proteins, including rice protein (RP), peanut protein isolate (PPI), soybean protein isolate (SPI), wheat gluten (WG), and zein. By analyzing the output signals obtained from TENGs through pairwise combination tests, we obtained a detailed *triboelectric series* of plant proteins according to their electrical properties. Then, by integrating the proteins with a tribo-electronegative dielectric layer, that is, a polylactic acid (PLA) film, we developed a highly flexible bio-TENG. We used the fabricated the bio-TENG as a biodegradable mulch film to construct a growth-promoting system that generates space electric fields to promote crop growth.

## 2. Material and Methods

*Fabrication of plant protein films (RP film, PPI film, SPI film, WG film, and zein film)*: The film-forming solutions were prepared by dispersing RP, PPI, SPI, and WG in distilled water (5 %, w/w) while being stirred. After 10 min of magnetic stirring, the protein precursor solution was then adjusted to pH 12 using a 1.0 M NaOH solution to fully dissolve the protein. In particular, zein was dispersed in a 70 % (v/v) ethanol solution without alkaline conditions. Then, glycerol was added to the five protein solutions at 30 % (w/w, dry basis) of the protein. Subsequently, these protein solutions were heated while being stirred at 65 °C for 30 min in a water bath, following a degassing process under vacuum for 10 min to remove air bubbles prior to casting. The final film-forming solutions were poured into circular Teflon molds (60 mm in diameter) and dried at 65 °C in the oven. The dried films were then peeled off and conditioned in a constant-temperature humidity chamber at 25 °C and 40 % RH before testing.

*Fabrication of conductive electrode*: 10 mg of bis(trifluoromethane)sulfonamide lithium salt was added to 1 mL of 1.5 % PEDOT:PSS aqueous solution for enhancing the conductivity of PEDOT:PSS ink. After 10 min of shaking to mix them evenly, the PEDOT:PSS ink was then gelled for a period of 30 min and finally coated onto the pretreated hydrophilic PLA film to form a homogeneous film at 60 °C.

*Amino acid analysis*: The protein powder was hydrolyzed by 6 M HCl at a temperature of 110 °C for 24 h, and the hydrolysates were analyzed using an amino acid analyzer (L-8900, Hitachi, Japan). The flow rate was 0.4 mL/min, and the column temperature was set to 55 °C. A group of standard amino acids was used as the markers, in which the detection wavelength of proline was 440 nm, and the detection wavelength of other amino acids was 570 nm. The content of each kind of amino acid was calculated using the standard curves and expressed as g/100 g of protein.

*Fabrication of bio-TENG (biodegradable mulch film) used in the application of cultivation of bok choi*: In actual use, the area of application of the mulch film is quite large; therefore, to meet the requirement for easy scaling up, the fabrication of bio-TENG is optimized. While in the former experiment, the final film-forming solutions were poured into circular Teflon molds (60 mm in diameter) and dried at 65 °C in the oven, herein, the film-forming solutions were blade-coated onto the conductive electrode (PEDOT:PSS-coated PLA) and dried at 65 °C. Subsequently, the biodegradable mulch film was fabricated via integration with the PLA. In the cultivation of bok choi, we connected the bio-TENG through a commercial bridge rectifier consisting of four diodes, which transformed the produced alternating current (AC) output into direct current (DC). Then, the positive side was suspended on top of the bok choi.

## 3. Results and Discussion

Regarding the characterization of plant protein films, we considered rice, wheat, soybeans, corn, and peanuts, which are major crops worldwide. Many by-products are obtained during the processing of grains and oils (**Fig. 1a**). These by-products are rarely used for deep processing and reuse given their low economic value. Plant protein, as one by-product, was recycled to prepare plant protein films, which were then integrated into the proposed bio-TENG. The plant protein films prepared by simple casting exhibited excellent flexibility and a level of transparency (**Fig. 1b**). The surface and cross-sectional morphologies of the five plant protein films were characterized using scanning electron microscopy (SEM). Each protein film has a relatively flat continuous structure and a uniform thickness of 100–200 μm (**Fig. S1**).

We used the elongation at break and tensile strength to investigate the mechanical properties of the prepared plant protein films. **Fig. S2** shows the stress–strain curves of the five plant protein films. The tensile strength in descending order was observed in the WG, zein, PPI, SPI, and RP films. In addition, the elongation at break in descending order was observed in the SPI, WG, PPI, RP, and zein films. From these results, we could conclude that the plant protein films have a certain mechanical strength, being suitable to constitute dielectric layers of bio-TENGs.

To study the exact positions of the five plant protein films in the triboelectric series, we constructed a simple TENG working in vertical contact–separation mode to measure the relative charging polarity of each plant protein according to the representative materials in the triboelectric series [28]. Since the thickness and relative humidity can influence the electrical output performance, we cast the same volume of protein film-forming precursor and tested the electrical output of the TENG at a fixed humidity. The schematic diagram in **Fig. 1c** illustrates the working principle of the TENG with the representative RP film and PDMS as a triboelectric pair. The contact–separation steps are labeled I–IV in **Fig. 1c** and **1d**. Initially, the two films are electrically neutral without charge. As the two films come into contact, charges transfer from one film to the other (state III in **Fig. 1c** and **1d**). During the release, external electrons of the two electrodes flow through the external load to achieve energy balance (state IV in **Fig. 1c** and **1d**). After the release, the top and bottom electrodes maintain equal charges of opposing polarities (state I in **Fig. 1c** and **1d**). Once the two films are brought into contact again, the electrostatic equilibrium is broken, driving the electrons to flow back (state II in **Fig. 1c** and **1d**). As shown in **Fig. 1d**, a negative current is observed as the RP film approaches the PDMS. Therefore, the external electron flow goes from the RP film to the PDMS, that is, the RP film is positively charged during contact with the PDMS. During the release, a positive current is produced, indicating that the transferred electrons flow back through the external load in the opposite direction.

Based on these experimental results, we constructed TENG combinations between each protein film in the triboelectric series and nine typical materials: (+) nylon, wool, silk, Al, paper, cellulose acetate, PET, PDMS, and PTFE (−). Using a white light interferometer, we observed no substantial difference in the surface roughness between the five plant protein films. This implies that their output performance depends mainly on the corresponding proteins (**Fig. 2a–e**). **Fig. 2a** shows the output current signal of the RP-film-enabled TENGs. The wool-RP-film-enabled TENG (top wool layer and bottom RP film) produced a negative current signal, indicating that the RP film was negatively charged upon contact with wool. In contrast, the silk-RP-, Al-RP-, cellulose-acetate-RP-, PET-RP-, PDMS-RP-, and PTFE-RP-film-enabled TENGs showed the opposite polarity of the output current signal, indicating that the RP film was positively charged upon contact in these TENGs. Therefore, the RP film was placed between wool and silk in the triboelectric series. Similarly, we investigated the triboelectric charging behaviors of the PPI, SPI, WG, and zein films. As shown in **Fig. 2b**, the output signal polarity of the PPI-film-enabled TENG varied between that of wool and silk, being at the same location in the triboelectric series as the RP-film-enabled TENG. SPI and WG showed polarity changes between the polarities of silk and Al, demonstrating that they were triboelectrically negative with respect to RP and PPI (**Fig. 2c** and **2d**). In addition, the triboelectric positivity of zein was the smallest compared with the other four plant proteins, being located between the polarities of Al and paper (**Fig. 2e**).

To determine the exact order of triboelectric positivity for plant proteins, we obtained the surface charges of different protein films while in contact with each other. **Fig. 2** shows that the RP film is positively charged when in contact with the other four films, and the PPI film maintains a positive charge upon contact with the SPI, WG, and zein films. Multiple repeated measurements of the charge polarity allowed us to determine the following triboelectric positivity order of the plant proteins: (+) RP, PPI, WG, SPI, and zein (−). We found uncertainty in detecting the charge polarity of WG and SPI when they are in contact. Thus, we assume that they have a similar ability to supply electrons, and some unknown complex factors may influence electron transfer.

To further clarify the positivity order of the plant proteins, the outputs of TENGs consisting of plant proteins and PDMS were compared. The plant protein films were prepared by casting the same volume of film-forming solution, and before measurement, the films were left in an incubator at a fixed temperature of 25 °C and 40 % relative humidity (RH) for 24 hours. **Fig. 2g** and **2h** show the open-circuit voltage and short-circuit current of the plant protein-PDMS-enabled TENGs under an oscillating power of 15 W at 10 Hz, respectively. RP exhibited the highest output peak-to-peak voltage of 20.3 V and output current of 1.26 µA followed by PPI, SPI, WG, and zein. Excluding the order of SPI and WG, the other proteins in the triboelectric series were consistent in their results of the charge polarity measurements.

The triboelectric charging mechanisms of the plant protein films were also explored. Considering that triboelectric positivity is related to the chemical structure, such as nitrogen-containing groups and oxygen functional groups [29], we used attenuated total reflectance (ATR) Fourier-transform infrared spectroscopy (FTIR) for protein conformation analysis to explore the role of functional groups in the triboelectric charging behavior of the plant protein films [30, 31]. We focused on the analysis of the major protein regions of amides A, B, I, II, and III (**Fig. 3a**). **Fig. 3b** shows the peak at approximately 3288 cm$^{-1}$, indicating the existence of amide A in all the plant proteins. This broad absorption band results from the overlapping of the OH and NH stretching vibrations of functional groups, and the positions of the curves corresponding to RP, PPI, SPI, WG, and zein gradually move toward lower frequencies. These redshifts indicate an increase in the amount of hydrogen-bonded NH with OH of water [32], and amides are apt to be exposed to water to form hydrogen bonds [33]. Thus, the frequency shift of amide A reflects the increase of protein–water coupling [34]. These results are consistent with the triboelectric positivity order of the five plant protein films. The characteristic absorption peaks at approximately 2963 and 2925 cm$^{-1}$ (**Fig. 3b**) are attributed to the CH$_2$ symmetric ($v_s$CH$_2$) and asymmetric ($v_{as}$CH$_2$) stretching vibration modes, respectively. However, these two absorption peaks in the amide B region show no difference in any of the protein films. The absorption peaks at 1633 cm$^{-1}$ representing the total hydrogen-bonded peptide groups in the amide I region, 1480–1580 cm$^{-1}$ representing NH and CN in the amide II region, and 1230 cm$^{-1}$ representing CO, CN, and NH absorption in the amide III region are shown in **Fig. 3c**. Neither the red nor the blue shift of the absorption peaks of these groups in the five proteins can be observed. In addition, a high sensitivity to small variations in hydrogen bonding patterns and molecular geometry is visible, making the amide I region useful for analyzing protein secondary structural composition and conformational changes [35, 36]. Thus, we obtained a typical deconvoluted FTIR spectrum focusing on the amide I region of the RP film. **Fig. 3d** shows six major characteristic peaks of the secondary structures (1619, 1633, 1647, 1661, 1675, and 1689 cm$^{-1}$). The difference in the secondary structure of these films can be obtained from the peak area and position of the deconvolution [36]. The positions of the characteristic peaks (**Fig. S3a**) are almost constant (±1 cm$^{-1}$), and the relative areas (**Fig. S3b**) of the peaks show small differences between the protein films, indicating that the films have a similar secondary structure. Overall, after investigating the characteristic regions (amides A, B, I, II, and III) of the five plant protein films, we can conclude that the degree of coupling with water molecules is an important factor in triboelectric positivity.

To further understand the influence of the chemical structure on the bio-TENG, we also used the recently developed atomic-force-microscope infrared spectroscopy (AFM-IR) to obtain basic information on surface-distributed functional groups. Through the infrared absorption spectrum of the five plant protein films, the approximate positions of four peaks with strong absorption: 1210, 1462, 1546, and 1670 cm$^{-1}$ could be roughly confirmed. As shown in **Fig. 4a** and **4b**, the AFM-IR map at approximately 1210 cm$^{-1}$ is ascribed to the asymmetric vibrations of C-O-C groups related to asymmetric CH$_3$ rocking vibrations, and the AFM-IR map at approximately 1462 cm$^{-1}$ represents the distribution of the CH$_2$ bending vibration of the protein. In addition, the amide II mode (approximately 1546 cm$^{-1}$) shown in **Fig. 4c** arose mainly from the combination of NH in-plane bending and CN stretching vibration, and the amide I vibration (approximately 1670 cm$^{-1}$) shown in **Fig. 4d** arose mainly in the form of C=O stretching vibration with minor contributions from the out-of-phase CN stretching vibration, CCN deformation, and NH in-plane bending. Through the RGB (red–green–blue) overlay of δ(CH$_2$), amide I, and amide II, we could determine basic information of the surface distribution of functional groups on the plant proteins. Other than the most distributed CH vibrations, the nitrogen-containing groups were evenly distributed on the surface (**Fig. 4e** and

**Fig. S4–S7**).

To evaluate the contributions of functional groups to the contact electrification of each plant protein film, we conducted an amino acid analysis to roughly estimate how the triboelectric positivity order is related to the chemical structure of the proteins (**Table S1**). Nitrogen-containing structures, that is, pyridine, amine, and amide groups, develop the greatest positive charge in various chemical structures [29]. Thus, we added the components of nitrogen-containing functional groups on the amino acid side chain of each protein film to consider the comprehensive effect of triboelectric positivity. As shown in **Fig. 4f**, the RP film had the highest proportion of nitrogen-containing groups followed by the PPI, zein, WG, and SPI films. The result of the zein film appeared different from that in the former detection, possibly due to three factors: 1) pH condition for protein film fabrication, as the zein film is distinctive; 2) many issues can influence the contact charge that occurs on protein films, such as the surface characteristics, experimental conditions, and contact nature. Although we aimed to maintain similar experimental conditions, there existed some deviations; 3) the integrated triboelectric charging behavior of protein was influenced by the combined action of multiple functional groups, such as nitrogen-containing groups and carboxyl groups, which produced the most obvious effects.

We also evaluated the effect of carboxyl groups on the triboelectric charging behavior of the protein films by roughly counting the carboxyl groups of the amino acid side chains of each protein film (**Fig. 4g**). Unlike the amino group, the carboxyl group easily produced a negative charge, impairing the ability of the protein film to donate electrons. The lowest carboxyl content was present in the RP film followed by the PPI, WG, SPI, and zein films, being consistent with the previous experimental results and further verifying the obtained triboelectric series.

After confirming the triboelectric positivity of the plant protein films, the bio-TENG was then characterized. The RP film was selected as the model for integration with a PLA film to construct a bio-TENG, and its operation principle was confirmed through simulations (COMSOL Multiphysics) based on the finite-element method (**Fig. S8**). **Fig. S9** shows that the top and bottom layers of the PLA film encapsulate and protect the protein film, and the top PLA film also acts as a negative dielectric. A poly(3,4-ethylenedioxythiophene)-poly(styrene sulfonate) (PEDOT:PSS) layer was coated onto the surface of the bottom PLA film to form a conductive electrode (sheet resistance, 649 ± 20 mΩ sq$^{-1}$). The SEM images and corresponding element mapping images of the electrode surface indicate a uniform PEDOT:PSS distribution (**Fig. S10**).

The operation force and frequency dependences of a bio-TENG with a contact area of approximately 7.065 cm$^2$ (i.e., circle of 30 mm in diameter) were also analyzed. **Fig. S11** shows that the outputs of the bio-TENG increase with the increase in the oscillating power of the vibration exciter, and herein, the oscillating power was proportional to the striking force. For the frequency response of the bio-TENG, we applied operation frequencies from 1 to 25 Hz at a fixed power, and the outputs increased with increasing frequency. This is because the accumulation of charges on the film surface increases the output voltage and current. Furthermore, the stability of the RP-film-enabled bio-TENG was evaluated by a repetitive contact test for approximately 40,000 cycles at an operating frequency of 10 Hz and a fixed oscillating power of 15 W. The results in **Fig. S12** show a stable voltage output over the cycles.

As the triboelectric layer is a protein film, two key factors may influence the bio-TENG output, namely, the RH and film thickness. As shown in **Fig. S13a**, at RP film thickness of approximately 51 μm, as RH increased from 40% to 70%, the output voltage decreased from 9.9 to 2.5 V, and the current decreased accordingly. When the RH increased further (i.e., 80% and 90%), the output began to increase. During contact electrification, either ion transfer or electron transfer can contribute to the charged surface of the dielectric layers, and water plays the most important role in the type of contribution. Specifically, as the RH varied from 40% to 70%, the TENG was dominated by electron transfer, and the presence of water on the triboelectric layers caused charges to leak away. In contrast, as the RH further increased to 80% or 90%, the water layer acted as a water bridge for ions to move and increase the charge density on the triboelectric layers, and the TENG was dominated by ion transfer. To evaluate thickness, we prepared four RP films with varying thicknesses using film-forming solutions of 4, 6, 8, and 10 mL poured into molds of the same size. We obtained approximate thicknesses of 51, 125, 187, and 431 μm, respectively (**Fig. S14**). **Fig. S13b** shows that the films created with 4 mL, 6 mL, and 8 mL solutions produced similar outputs at an RH of 90%, whereas that created with the 10 mL solution exhibited inferior performance. The triboelectric charge density has been related to the capacitance of the bio-TENG, surface area, thickness, and relative dielectric constant [37]. Theoretically, a thinner film should produce a higher output. However, thicker protein films result in more adsorbed water on the account of more glycerol. Therefore, mobile ions can transfer to enhance the charge density on the triboelectric layers. Considering these two factors, the RP films created with 4 mL, 6 mL, and 8 mL solutions generated similar outputs, whereas the film created with the 10 mL solution resulted in a high thickness that reduced the output performance.

Then, the biodegradability of the proposed bio-TENG was evaluated. For the degradation test, we buried the RP film and PEDOT:PSS-coated PLA film in soil. During degradation, the appearance and weight loss of the RP films were recorded. We observed that the RP film degraded more easily than the PEDOT:PSS-coated PLA film, and the weight over time in **Fig. S15a** indicates that the RP film completely degraded after 127 days. We also performed tensile strength tests to analyze the degradation properties of the PEDOT:PSS-coated PLA film. The mechanical strength of the composite film decreased after 143 days (**Fig. S15b**), indicating structural damage. These results demonstrate that the RP-film-enabled bio-TENG is fully biodegradable and can be recycled by the environment.

Bio-TENGs can be used as biodegradable mulch films that allow the creation of growth-promoting systems. Given the pollution generated by residual plastic mulch in soil, biodegradable mulch films have become an alternative solution [38], and PLA is an environmentally friendly polymer material that has attracted interest in film mulching [39]. Accordingly, we created a PLA-packaged bio-TENG with full biodegradability, high flexibility, good mechanical properties, and outstanding stability as a biodegradable mulch film. Considering the intrinsic voltage generation capability, the RP-film-enclosed bio-TENG can produce a space electric field to form a growth-promoting system for agriculture. We used hand slapping to simulate people walking on a mulch film. **Fig. S16** shows that force exerted on the biodegradable mulch film (*i.e.*, PLA-packaged bio-TENG) changed the distance between the PLA and RP films, generating a voltage signal, as illustrated in **Movie S1**. We used bean seedlings as a model to study growth with and without applying an electric field. The electric field was generated by the fabricated bio-TENG through an exciter system, whose diagram is shown in **Fig. 5a**. The two groups of bean seedlings grew in identical conditions (25 °C, 40% RH), but the test group was subjected to a 1 h electric field every 12 h. When the mechanical excitation on the bio-TENG was applied to simulate the mulch harvesting mechanical energy, the output peak-to-peak voltage reached approximately 180 V (contact area: 28.26 $cm^2$; operating frequency: 15 Hz; oscillating power: 15 W), as shown in **Fig. S17**. **Fig. 5b** and **5c** show that the bean seedlings of test group grew better than those of the control group with regard to length and weight. The growth in the test group (average length and weight growth rates of 223.8% and 24.3%, respectively) was significantly better (****$p < 0.0001$) than that in the control group (average length and weight growth rates of 132.4% and 19.1%). The increase in seed germination and seedling growth may be related to water absorption by the seeds [40, 41]. In addition, we conducted an experiment on mung beans, finding that the weight of mung bean seeds after applying the electric field treatment was significantly greater than that of the control group (****$p < 0.0001$). This result indicates that electric field treatment improves the water absorption rate of the mung bean seeds (**Fig. S18**) and indirectly improves seed germination. Therefore, we confirmed the feasibility of using bio-TENGs to generate a space electric field to form a growth-promoting system for agriculture (**Fig. 5d**).

To demonstrate the applicability of bio-TENGs in agriculture, the growth of bok choi in greenhouses was monitored over time. Four groups of bok choi with similar growth, each of which comprised 20 bok choi seedlings, were used to evaluate the effect of bio-TENGs as mulch film on crop growth. We recorded three parameters, namely, plant height, crown diameter, and the number of leaves, to determine the state of growth. **Fig. 6** shows the obtained results expressed as mean ± standard deviation ($n = 20$). Group 1 was the experimental group with the bio-TENGs used as mulch film. Group 2 was planted next to group 1, and groups 3 and 4 were planted farther away (compared to group 2) from group 1 (**Fig. S19**).

**Fig. 6a**, **6c**, and **6e** show that the plant height, crown diameter, and the number of leaves changed significantly over time (****$p < 0.0001$) compared with the original state. Moreover, we compared the growth conditions across groups over time. **Fig. 6b**, **6d**, and **6f** show no initial difference in plant height, crown diameter, and the number of leaves between the four groups. However, compared with groups 3 and 4, the experimental group (group 1) and group 2 (close to group 1) grew better over time. Regarding plant height, on the 6$^{th}$ day, groups 1 and 2 were similar, while groups 3 (****$p < 0.0001$) and 4 (*$p < 0.05$) exhibited differences compared with group 1. This trend became more obvious over time. On the 11$^{th}$ day, there was still no significant difference between groups 1 and 2, but the plant heights of groups 3 (***$p < 0.001$) and 4 (*$p < 0.05$) were visibly lower than those of group 1. Regarding crown diameter, on the 6$^{th}$ day, group 2 was similar to group 1, while group 3 and group 4 showed impaired crown diameter as compared with group 1 (****$p < 0.0001$). However, on the 11$^{th}$ day, groups 3 and 4 (****$p < 0.0001$) showed a lower crown diameter than group 2, whose crown diameter was lower than that of group 1 (**$p < 0.01$). Regarding the number of leaves, on the 6$^{th}$ day, groups 1, 2, and 3 were similar, whereas group 4 showed a significant difference (**$p < 0.01$). On the 11$^{th}$ day, besides group 4 (****$p < 0.0001$), there was also a difference between groups 3 and 1 (****$p < 0.0001$).

The growth of bok choi can be seen in photographs. We show the original state in **Fig.**

**S19a** and **S19b**. On the 6[th] and 11[th] days, groups 1 and 2 (**Fig. S19d** and **S19f**) grew better than groups 3 and 4 (**Fig. S19c** and **S19e**). We finally measured the weight of each bok choi at harvest time. Although their original weights were unknown, the final weight data could provide a rough reference for growth. **Fig. S20** shows a similar growth trend as the abovementioned results. Overall, our results suggest that using bio-TENGs as mulch films can promote the growth of bok choi, and plants closer to the electric field produced by the mulch film exhibit more obvious growth effects. Plant-protein-enabled bio-TENGs seem promising to enhance crop yields, possibly establishing a new generation of mulch films that harvest mechanical energy from the environment to produce electrical energy that stimulates crop growth (**Fig. 5d**).

## 4. Conclusions

In this paper, we introduced five representative plant proteins (RP, PPI, SPI, WG, and zein) as bio-TENG positively charging materials. The triboelectric series of these representative plant proteins were ranked in descending order of positivity as RP followed by PPI, SPI, WG, and zein. After exploring the bio-TENG mechanism using ATR-FTIR, AFM-IR, and amino acid analysis, we predicted that the triboelectric positivity of the plant protein films is determined by multiple chemical structures and factors, such as nitrogen-containing groups, carboxyl groups, and the degree of coupling with water molecules, establishing theoretical foundations for the proposed bio-TENG. Furthermore, we used plant-protein-enabled bio-TENGs as a new type of mulch film to construct a growth-promoting system that generates space electric fields. After verifying the faster growth rate of bean seedlings, the significantly better growth of greenhouse bok choi cultivated with bio-TENG mulch film in a real agricultural environment further confirmed the applicability of the growth-promoting system. From this, we can conclude that plant proteins are new triboelectric materials that can replace non-biodegradable synthetic polymers and broaden the material selection for bio-TENGs. Adopting such bio-TENGS may contribute to solving problems related to electronic waste, residual plastic of mulch films, and waste of plant protein sources, etc., suggesting their potential for application in sustainable agriculture.

**Declaration of Competing Interest**

The authors declare no conflict of interest in this work.

**Acknowledgments**

This research was supported by the National Natural Science Foundation of China for Excellent Young Scholars (Grant No. 31922063).**References**

[1] X. Wang, Y. Zhang, X. Zhang, et al., A Highly Stretchable Transparent Self-Powered Triboelectric Tactile Sensor with Metallized Nanofibers for Wearable Electronics, Adv. Mater. 30 (2018) 1706738.

[2] Y. Lee, S.H. Cha, Y.W. Kim, et al., Transparent and attachable ionic communicators based on self-cleanable triboelectric nanogenerators, Nat. Commun. 9 (2018) 1804.

[3] C. Wu, A.C. Wang, W. Ding, et al., Triboelectric Nanogenerator: A Foundation of the Energy for the New Era, Adv. Energy Mater. 9 (2019) 1802906.

[4] R. Ccorahua, J. Huaroto, C. Luyo, et al., Enhanced-performance bio-triboelectric nanogenerator based on starch polymer electrolyte obtained by a cleanroom-free processing method, Nano Energy 59 (2019) 610-618.

[5] G. Wei, Y. Bi, X. Li, et al., Self-powered hybrid flexible nanogenerator and its application in bionic micro aerial vehicles, Nano Energy 54 (2018) 10-16.

[6] Q. Zhang, Q. Liang, Q. Liao, et al., An Amphiphobic Hydraulic Triboelectric Nanogenerator for a Self-Cleaning and Self-Charging Power System, Adv. Funct. Mater. 28 (2018) 1803117.

[7] P. Cheng, Y. Liu, Z. Wen, et al., Atmospheric pressure difference driven triboelectric nanogenerator for efficiently harvesting ocean wave energy, Nano Energy 54 (2018) 156-162.

[8] C. Jiang, X. Li, Y. Yao, et al., A multifunctional and highly flexible triboelectric nanogenerator based on MXene-enabled porous film integrated with laser-induced graphene electrode, Nano Energy 66 (2019) 104121.

[9] L. Lan, T. Yin, C. Jiang, et al., Highly conductive 1D-2D composite film for skin-mountable strain sensor and stretchable triboelectric nanogenerator, Nano Energy 62 (2019) 319-328.

[10] Y. Dong, S.S.K. Mallineni, K. Maleski, et al., Metallic MXenes: A new family of materials for flexible triboelectric nanogenerators, Nano Energy 44 (2018) 103-110.

[11] D.W. Kim, J.H. Lee, I. You, et al., Adding a stretchable deep-trap interlayer for high-performance stretchable triboelectric nanogenerators, Nano Energy 50 (2018) 192-200.

[12] C. Wu, T.W. Kim, J.H. Park, et al., Enhanced Triboelectric Nanogenerators Based on $MoS_2$ Monolayer Nanocomposites Acting as Electron-Acceptor Layers, ACS Nano 11 (2017) 8356-8363.

[13] X. Gao, L. Huang, B. Wang, et al., Natural Materials Assembled, Biodegradable, and Transparent Paper-Based Electret Nanogenerator, ACS Appl. Mater. Interfaces 8 (2016) 35587-35592.

[14] Q. Zheng, Y. Zou, Y. Zhang, et al., Biodegradable triboelectric nanogenerator as a life-time designed implantable power source, Sci. Adv. 2 (2016) e1501478.

[15] R. Wang, S. Gao, Z. Yang, et al., Engineered and Laser-Processed Chitosan Biopolymers for Sustainable and Biodegradable Triboelectric Power Generation, Adv. Mater. 30 (2018) 1706267.

[16] H.-J. Kim, J.-H. Kim, K.-W. Jun, et al., Silk Nanofiber-Networked Bio-Triboelectric Generator: Silk Bio-TEG, Adv. Energy Mater. 6 (2016) 1502329.

[17] C. Jiang, C. Wu, X. Li, et al., All-electrospun flexible triboelectric nanogenerator based on metallic MXene nanosheets, Nano Energy 59 (2019) 268-276.

[18] R. Pan, W. Xuan, J. Chen, et al., Fully biodegradable triboelectric nanogenerators based on electrospun polylactic acid and nanostructured gelatin films, Nano Energy 45 (2018) 193-202.

[19] H.-J. Kim, E.-C. Yim, J.-H. Kim, et al., Bacterial Nano-Cellulose Triboelectric Nanogenerator, Nano Energy 33 (2017) 130-137.

[20] M.N. Nasrabadi, A.S. Doost, R. Mezzenga, Modification approaches of plant-based proteins to improve their techno-functionality and use in food products, Food Hydrocoll. 118 (2021) 106789.

[21] A.J. Capezza, W.R. Newson, R.T. Olsson, et al., Advances in the use of protein-based materials: toward sustainable naturally sourced absorbent materials, ACS Sustain. Chem. Eng. 7 (2019) 4532-4547.

[22] T. Wang, X. Chen, Q. Zhong, et al., Facile and efficient construction of water-soluble biomaterials with tunable mesoscopic structures using all-natural edible proteins, Adv. Funct. Mater. 29 (2019) 1901830.

[23] D. Dannehl, Effects of electricity on plant responses, Sci. Hortic. 234 (2018) 382-392.

[24] D. Dannehl, S. Huyskens-Keil, D. Wendorf, et al., Influence of intermittent-direct-electric-current (IDC) on phytochemical compounds in garden cress during growth, Food Chem. 131 (2012) 239-246.

[25] D. Dannehl, S. Huyskens-keil, I. Eichholz, et al., Effects of direct-electric-current on secondary plant compounds and antioxidant activity in harvested tomato fruits (Solanum lycopersicon L.), Food Chem. 126 (2011) 157-165.

[26] M. Guderjan, S. Topfl, A. Angersbach, et al., Impact of pulsed electric field treatment on the recovery and quality of plant oils, J. Food Eng. 67 (2005) 281-287.

[27] T. Fan, S. Wang, Y. Li, et al., Film mulched furrow-ridge water harvesting planting improves agronomic productivity and water use efficiency in Rainfed Areas, Agric. Water Manag. 217 (2019) 1-10.

[28] M. Seol, S. Kim, Y. Cho, et al., Triboelectric Series of 2D Layered Materials, Adv. Mater. 30 (2018) e1801210.

[29] A.F. Diaz, R.M. Felix-Navarro, A semi-quantitative tribo-electric series for polymeric materials: the influence of chemical structure and properties, J. Electrostat. 62 (2004) 277-290.

[30] S.K. Ghosh, D. Mandal, Efficient natural piezoelectric nanogenerator: Electricity generation from fish swim bladder, Nano Energy 28 (2016) 356-365.

[31] S.B. Akkas, M. Severcan, O. Yilmaz, et al., Effects of lipoic acid supplementation on rat brain tissue: An FTIR spectroscopic and neural network study, Food Chem. 105 (2007) 1281-1288.

[32] C.H. Bamford, L. Brown, A. Elliott, et al., Structure of synthetic polypeptides, Nature 169 (1952) 357-358.

[33] W. Gallagher, FTIR analysis of protein structure, Course manual Chem 455 (2009).

[34] F. Demmel, W. Doster, W. Petry, et al., Vibrational frequency shifts as a probe of hydrogen bonds: thermal expansion and glass transition of myoglobin in mixed solvents, Eur. Biophys. J. 26 (1997) 327-335.

[35] H.Y. Yang, S.N. Yang, J.L. Kong, et al., Obtaining information about protein secondary structures in aqueous solution using Fourier transform IR spectroscopy, Nat. Protoc. 10 (2015) 382–396.

[36] J. Kong, S. Yu, Fourier transform infrared spectroscopic analysis of protein secondary structures, Acta Biochim. Biophys. Sin. 39 (2007) 549-559.

[37] V. Harnchana, H.V. Ngoc, W. He, et al., Enhanced power output of a triboelectric nanogenerator using poly (dimethylsiloxane) modified with graphene oxide and sodium dodecyl sulfate, ACS Appl. Mater. Interfaces 10 (2018) 25263-25272.

[38] S. Kasirajan, M. Ngouajio, Polyethylene and biodegradable mulches for agricultural applications: a review, Agron. Sustain. Dev. 32 (2012) 501-529.

[39] I. Kwiecien, G. Adamus, G.Z. Jiang, et al., Biodegradable PBAT/PLA Blend with Bioactive MCPA-PHBV Conjugate Suppresses Weed Growth, Biomacromolecules 19 (2018) 511-520.

[40] B. Sera, V. Stranak, M. Sery, et al., Germination of Chenopodium album in response to microwave plasma treatment, Plasma Science & Technology 10 (2008) 506-511.

[41] H.H. Chen, Y.K. Chen, H.C. Chang, Evaluation of physicochemical properties of plasma treated brown rice, Food Chem. 135 (2012) 74-79.

**Figures**

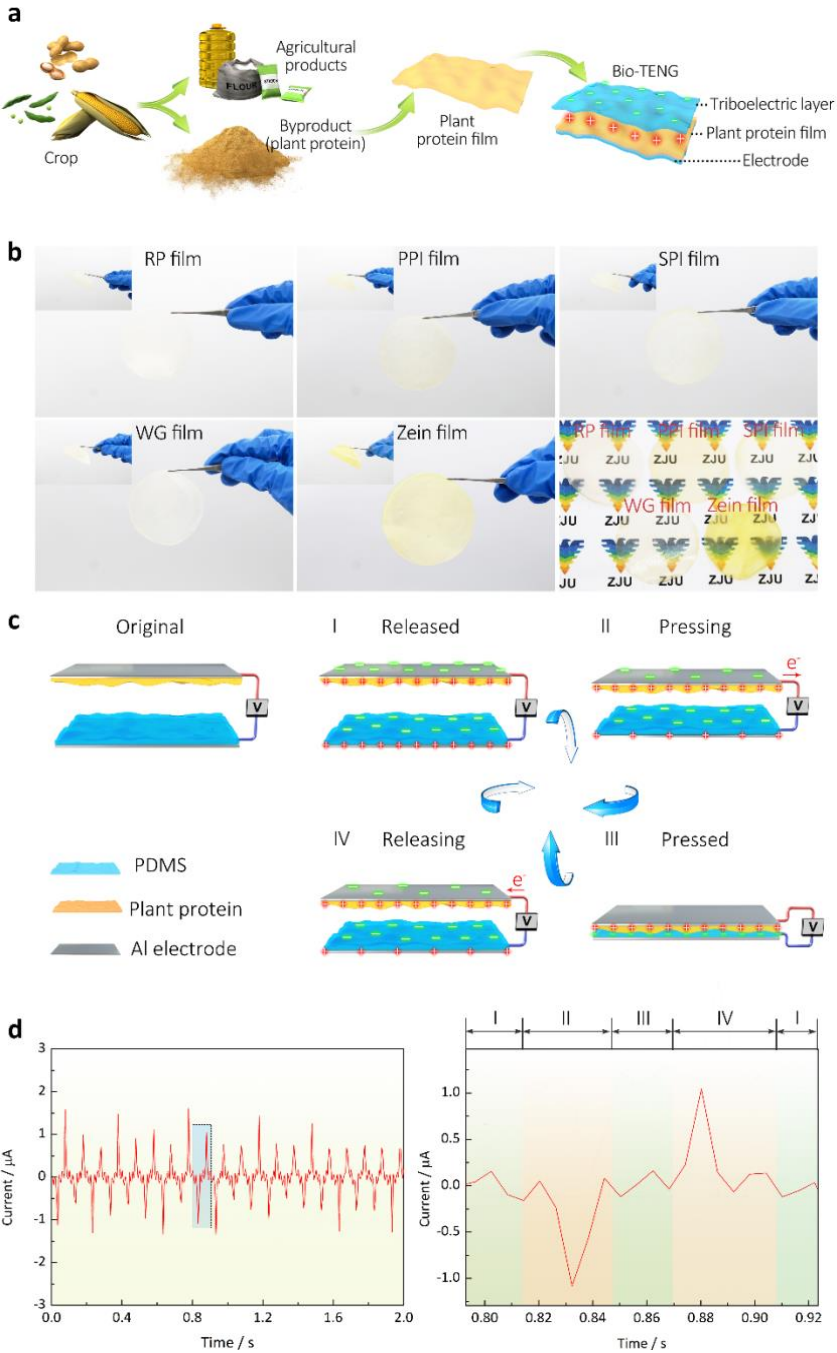

**Fig. 1.** Triboelectric charging behaviors and output current signal. (a) Fabrication of plant-protein-film-enabled bio-TENG. (b) Photographs of RP, PPI, SPI, WG, and zein films. The plant protein films have some level of transparency. (c) Schematic diagram showing triboelectric charging behavior of plant protein film (using Al as conductive electrode and PDMS as triboelectric negative layer). (d) Output current signal and its magnified view over one cycle

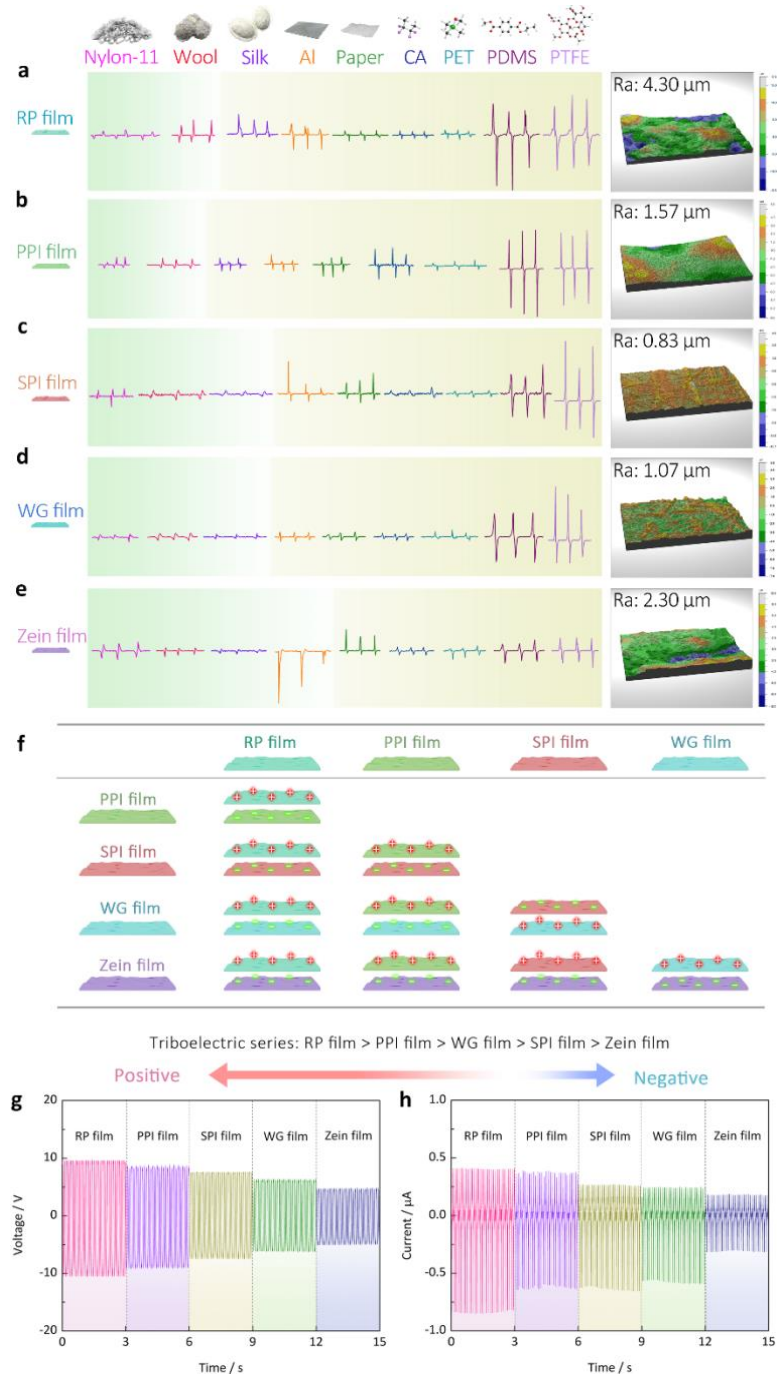

**Fig. 2.** Triboelectric charging behavior of various plant protein films (CA, cellulose acetate). Output current signals of TENGs with (a) RP, (b) PPI, (c) SPI, (d) WG, and (e) zein as triboelectric materials and surface roughness of the corresponding plant protein films. (f) Charging of plant protein films upon contact with another film. (g) Open-circuit voltage and (h) short-circuit current of plant-protein-PLA-enabled bio-TENGs (plant protein film as triboelectric positive layer and PLA as triboelectric negative layer).

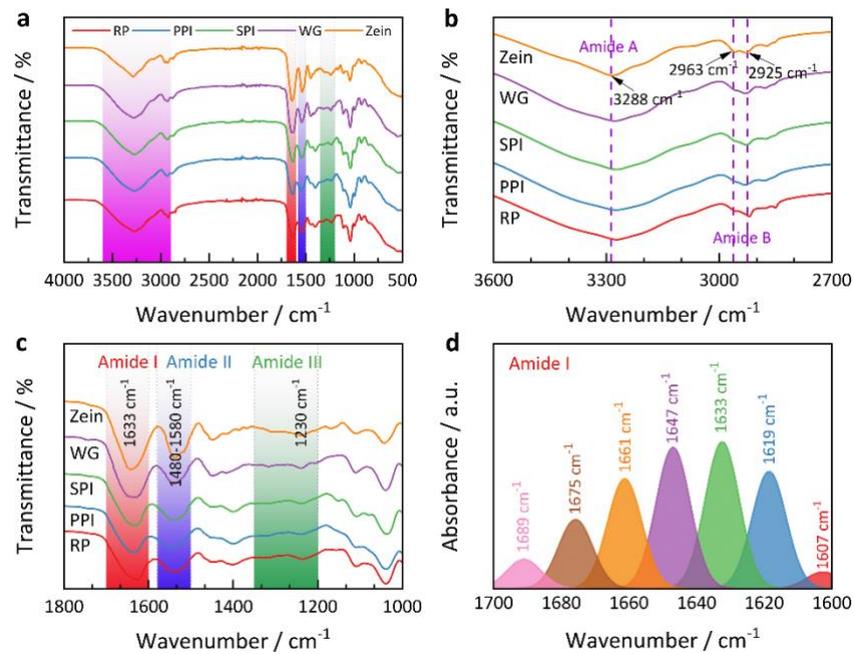

**Fig. 3.** ATR-FTIR analysis of five plant protein films. ATR-FTIR spectra of plant protein films in (a) 4000–500 cm$^{-1}$ and (b) 3600–2700 cm$^{-1}$. (c) ATR-FTIR spectra in 1800–1000 cm$^{-1}$. (d) Gaussian-curve-fit inverted second-derivative amide I spectra of RP film.

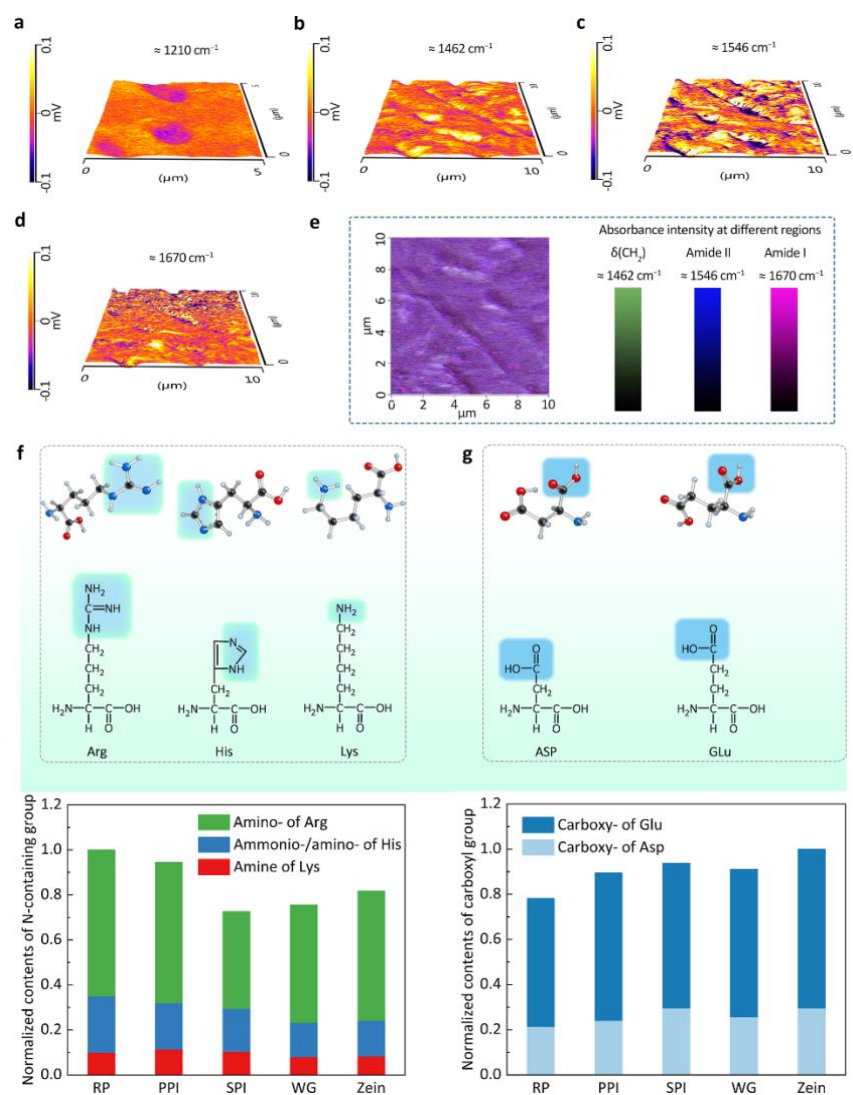

**Fig. 4.** AFM-IR map of RP film at (a) 1210, (b) 1462, (c) 1546, and (d) 1670 cm$^{-1}$. (e) RGB overlay of regions at 1462, 1546, and 1670 cm$^{-1}$ representing δ(CH$_2$), amide II, and amide I, respectively. Stick model and molecular formula of amino acids including (f) nitrogen-containing and (g) carboxyl groups on the side chain and a normalized number of nitrogen-containing groups and carboxyl groups calculated by amino acid analysis of five plant protein films. (Arg, arginine; His, histidine; Lys, lysine; Asp, aspartic acid; Glu, glutamic acid)

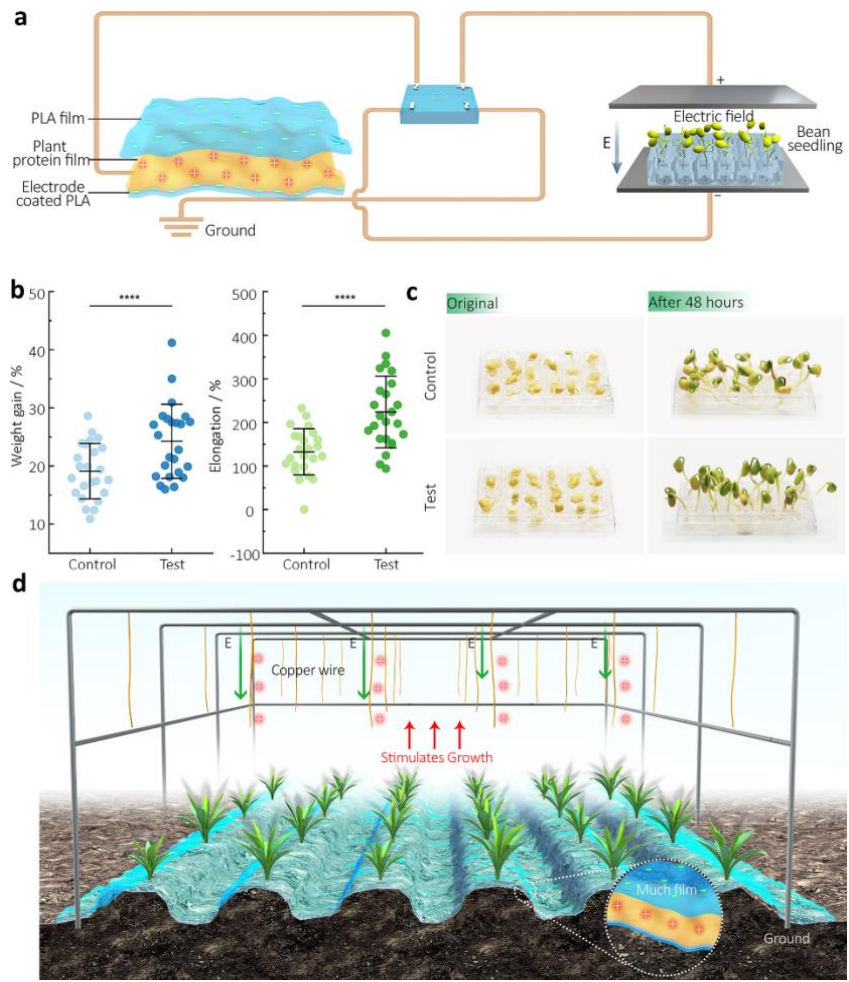

**Fig. 5.** Bio-TENG as a biodegradable mulch film to construct a growth-promoting system that generates space electric fields. (a) Schematic diagram of the electric field experiment. (b) Percentage of weight gain (left) and elongation (right) in control (without electric field) and test (with electric field) groups after 48 h of growth. Whiskers indicate mean ± standard deviation for $n$ = 24 bean seeds per group. Two-sample $t$-test; **** $p < 0.0001$. (c) Typical photographs of beans with (test) and without (control) electric field application before and after 48 h of growth. (d) Plant-protein-enabled bio-TENG used as biodegradable mulch film to construct growth-promoting system that generates space electric field for agriculture.

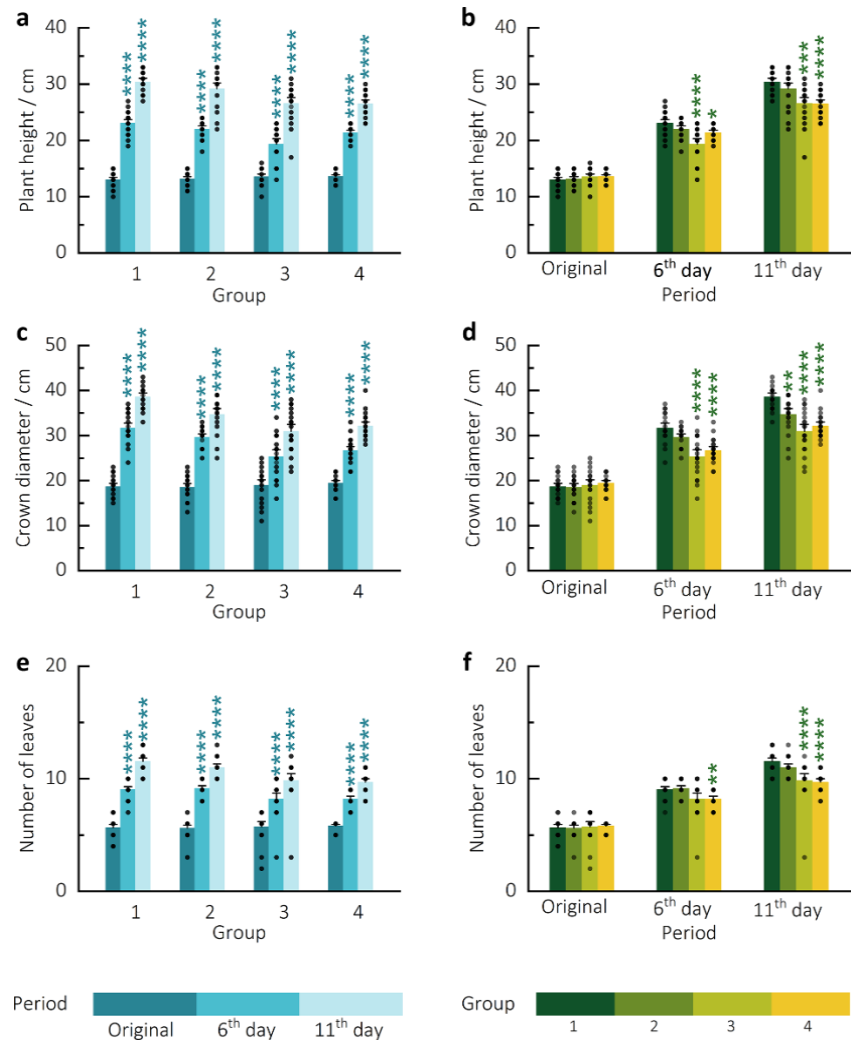

**Fig. 6.** Bio-TENG as a biodegradable mulch film to construct growth-promoting system that generates space electric field for bok choi crops. Data are expressed as mean ± standard deviation for *n* = 20 bok choi seedlings per group analyzed by one-way ANOVA (analysis of variance); *$p$ < 0.05; **$p$ < 0.01; ***$p$ < 0.001; ****$p$ < 0.0001. (a) Plant height of four bok choi groups over time and (b) difference in plant height between groups in different periods. (c) Crown diameter over time of four bok choi groups and (d) difference in crown diameter between groups in different periods. (e) Number of leaves over time of four bok choi groups and (f) difference in the number of leaves between groups in different periods.